\documentclass{PoS}
\usepackage{bm}

\title{Top-quark hadroproduction in NNLO QCD}

\ShortTitle{Top-quark hadroproduction in NNLO QCD}

\author{\speaker{Simone Devoto}\,\footnote{Work done in collaboration with Stefano Catani, Massimiliano Grazzini, Stefan Kallweit and Javier Mazzitelli.}\\
        Physik Institut, Universit\"at Z\"urich, CH-8057 Z\"urich, Switzerland\\
        E-mail: \email{devoto@physik.uzh.ch}}

\abstract{The study of top-quark production and decay is central in the LHC physics programme, allowing precise tests of the Standard Model and offering a window on possible new physics. Accurate theoretical predictions for this process are therefore crucial. We report on a new calculation of the next-to-next-to-leading order QCD radiative corrections to the production of top-quark pairs at hadron colliders. The calculation is performed by using the $q_T$-subtraction method to handle and cancel infrared singular contributions at intermediate stages of the computation, and it represents its first complete application to the hadroproduction of a colourful high-mass system at next-to-next-to-leading order. We discuss the calculation of the additional soft contributions needed to implement $q_T$ subtraction for this process, and we show first numerical results.}

\FullConference{14th International Symposium on Radiative Corrections (RADCOR2019)\\
9-13 September 2019\\
		Palais des Papes, Avignon, France}

\begin{document}

\section{Introduction}
Top quark ($t$) production is of great importance for physics at hadron colliders, and thus, in particular, for the LHC physics programme.
The high value of its mass, and consequently its strong coupling with the Higgs boson, makes the top quark relevant for electroweak symmetry breaking.
Besides, top production also plays an important role in searches for physics Beyond the Standard Model, since it constitutes a possible window on new physics and, at the same time, a crucial background in many analyses.

The main source of top-quark events in hadronic collisions is top-quark pair production.
Next-to-leading order (NLO) QCD corrections to this process have been available for thirty years, both for total cross sections and differential distributions \cite{Nason:1987xz, Beenakker:1988bq, Beenakker:1990maa, Nason:1989zy,Mangano:1991jk}.
A few years ago, also next-to-next-to-leading order (NNLO) QCD corrections were computed and results were again provided both for the total cross section~\cite{Baernreuther:2012ws, Czakon:2012zr, Czakon:2012pz, Czakon:2013goa} and for differential distributions~\cite{Czakon:2015owf,Czakon:2016ckf,Czakon:2017dip}. Further progress regards the combination of QCD and EW corrections~\cite{Czakon:2017wor} and the inclusion of top decays~\cite{Behring:2019iiv}.

In the following, we will focus on a new computation of NNLO QCD corrections we recently performed by using the $q_T$-subtraction formalism \cite{Catani:2019iny, Catani:2019hip}.
There are various reasons that motivated us to carry out this work. The first was to provide a first independent check of such an involved calculation. Besides, our results represent the first complete application of the $q_T$-subtraction formalism to the hadroproduction of a colourful final state at NNLO QCD accuracy.
Finally, there is no NNLO generator for differential distributions publicly available yet for this process, and our calculation is a first step for the inclusion of top pair production in a future public {\sc Matrix} release.\footnote{{\sc Matrix}~\cite{Grazzini:2017mhc} is a public code that, by using $q_T$ subtraction, allows the evaluation of fully differential cross sections at NNLO QCD for a wide class of processes in which a colourless final state is produced in hadronic collisions.}

\section{$\boldsymbol{q_T}$-subtraction formalism for heavy quark production}
At the parton level, the ingredients needed for the NNLO corrections to the $t\bar t$ cross section are tree-level contributions with two additional partons in the final state, one-loop amplitudes with an additional emitted parton and purely virtual contributions, which were all already known.
Nevertheless, the combination of the above contributions to construct a complete NNLO computation is a non-trivial task because of the presence of infrared (IR) divergences at intermediate steps of the calculation that do not allow a straightforward implementation of numerical techniques.

While at NLO the problem can nowadays be considered solved, with Catani-Seymour dipole subtraction \cite{Catani:1996jh,Catani:1996vz,Catani:2002hc} and FKS subtraction \cite{Frixione_1996} being the most widely employed methods to overcome these difficulties, at NNLO different methods have been proposed to handle and cancel IR singularities (see e.g. \cite{Bendavid:2018nar} and references therein), but none can yet claim full generality.

Our procedure is based on the $q_T$-subtraction formalism \cite{Catani:2007vq}:
originally developed for colourless final state, the $q_T$-subtraction formalism uses the knowledge we have from resummation studies of the low-$q_T$ behaviour
to construct a counterterm.
Considering a process where a colourless final state $F$ is produced, \mbox{$pp\to F  + X$}, the starting point is to split its differential NNLO cross section $d\sigma^F_{NNLO}$ into a part with $q_T=0$ and one with $q_T\neq 0$, $q_T$ being the transverse momentum of the final state $F$,
\begin{equation}
  d\sigma^F_{NNLO}=d\sigma^F_{NNLO}\big|_{q_T=0}+d\sigma^F_{NNLO}\big|_{q_T\neq 0}\;.
\end{equation}
Since at Born level the final state $F$ has $q_T=0$, the NNLO contributions at $q_T\neq 0$ are actually given by NLO contributions for the final state $F$+jets. One can hence write the cross section as follows,
\begin{equation}
  \label{eq:qt_sub}
  d\sigma^F_{NNLO}=\mathcal H^F_{(N)NLO}\otimes d\sigma_{LO}^F+\left[d\sigma_{NLO}^{F+jets}-d\sigma^{CT}_{NLO}\right]\;.
\end{equation}
The contribution at $q_T\neq0$ is now $d\sigma_{NLO}^{F+jets}$ that, being a NLO contribution, can be treated with standard NLO techniques.
Extra singularities of the NNLO type, associated with the \mbox{$q_T\to0$} limit, still need additional subtraction: this is achieved by introducing the counterterm $d\sigma^{CT}_{NLO}$, which has been derived by using the knowledge of the IR behaviour one has from the $q_T$-resummation formalism~\cite{Zhu:2012ts,Li:2013mia,Catani:2014qha}.
Finally, the information on the virtual corrections to the process, containing the $q_T=0$ contribution, is embodied in the coefficient $\mathcal H^F_{(N)NLO}$.

With the inclusion of additional contributions, the $q_T$-subtraction formalism can be extended to the case of a colourful final state. This had already been achieved for top pair production at NLO and at NNLO considering only off-diagonal channels~\cite{Bonciani:2015sha}. The NNLO computation of the diagonal channels has now been completed~\cite{Catani:2019iny,Catani:2019hip}.

The generalisation to a colourful final state involves the treatment of the additional final-state soft singularities. This does not require extra efforts for the computation of $d\sigma_{NLO}^{t\bar t+jets}$, which is computable with NLO subtraction techniques, while it requires a generalisation of the counterterm $d\sigma^{CT}_{NLO}$ and the coefficient $\mathcal H^{t\bar t}_{(N)NLO}$.
The expression of the counterterm is obtained by performing a perturbative expansion~\cite{Bozzi:2005wk,Bozzi:2007pn,Bonciani:2015sha} of the resummation formula  at NNLO for the $q_T$ distribution of the $t \bar t$ pair~\cite{Zhu:2012ts,Li:2013mia,Catani:2014qha}, and was thus already known.
The coefficient $\mathcal{H}^{t\bar t}_{NNLO}$ can be split into a process-dependent and a process-independent part:
its process-dependent part can be computed from the two-loop scattering amplitude~\cite{Czakon:2008zk,Baernreuther:2013caa}, while the process-independent part was already known for the colourless case, but not available when considering a colourful final state.
The difference between the former and the latter is of purely soft origin: while in the colourless case one has to consider only the collinear and soft radiation emitted by the massless initial-state partons, in the colourful case one also has to deal with soft radiation emitted off the massive final-state quarks.
The last missing ingredient to be taken care of was the integration of a suitably subtracted soft current, considering both the single gluon emission at one loop level~\cite{Catani_2000,Bierenbaum2012, Czakon2018} and the double-real emission~\cite{Catani1999, Czakon2011}.

We recently completed the computation of all the required integrals~\cite{inprep}: most of them have been performed analytically, whereas a few of the most challenging ones were evaluated numerically.\footnote{It is worth observing that, while we were progressing with our analytic computations, a fully numerical calculation has been carried out in the SCET formalism~\cite{Angeles-Martinez2018}.} With the inclusion of those contributions, all the ingredients needed for the generalisation of $q_T$ subtraction to top pair production were available.

\section{Numerical results}
The integration of the missing soft contributions allowed us to implement top pair production in the {\sc Matrix} framework~\cite{Grazzini:2017mhc}, and to compute inclusive cross sections~\cite{Catani:2019iny} as well as single- and double-differential distributions~\cite{Catani:2019hip} for this process.

The core of {\sc Matrix} is the Monte Carlo program {\sc Munich}\footnote{{\sc Munich} is the abbreviation of "MUlti-chaNnel Integrator at Swiss (CH) precision", an automated parton-level NLO generator by S. Kallweit.}, which includes an implementation of the dipole-subtraction method for NLO computations with massive final states~\cite{Cascioli:2013wga} and an efficient phase space integration. All the required amplitudes are obtained from {\sc OpenLoops 2}~\cite{Buccioni:2019sur}, except for the four-parton tree-level colour correlators that rely on an analytic implementation, and for the two-loop amplitude that relies on the numerical grid provided in Ref.~\cite{Czakon:2008zk,Baernreuther:2013caa}. In order to validate our results for the real--virtual corrections, we also employed the matrix-element generator {\sc Recola}~\cite{Actis:2016mpe,Denner:2017wsf} and found perfect agreement. The subtraction in the square bracket of Eq.(\ref{eq:qt_sub}) is, in practice, not implemented locally, but by introducing a cut-off $r_{cut}$ on the variable $r=q_T/M$, $M$ being the invariant mass of the top pair. The final result is obtained by performing the limit $r_{cut}\to 0$.

\begin{figure}[t]
  \centering
  \includegraphics[width=0.5\textwidth]{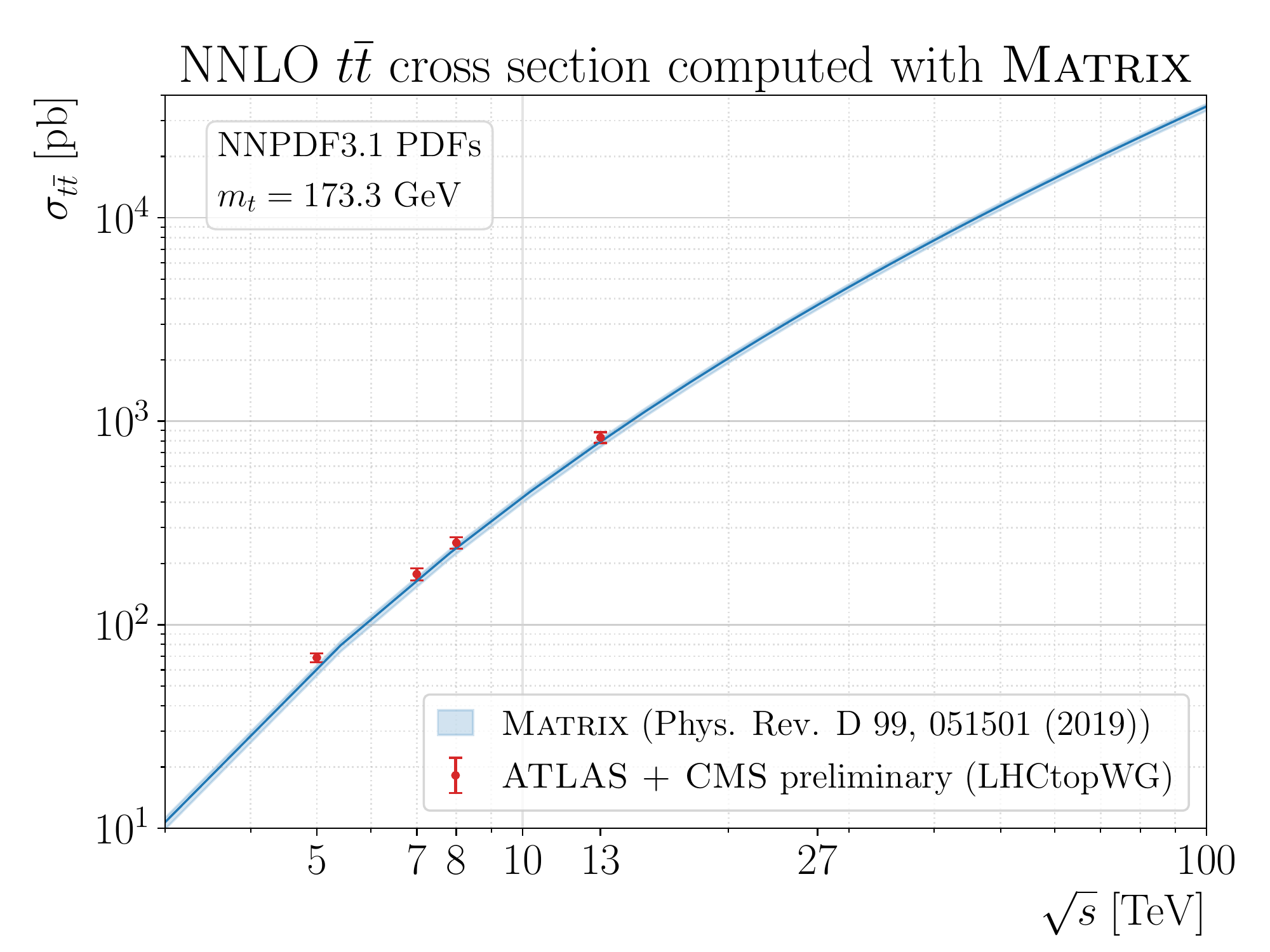}
  \vspace{-0.5cm}
  \caption{Inclusive cross section as a function of the collider energy.}
  \label{fig:inclusive}
\end{figure}

We start by presenting the inclusive cross section. In Figure \ref{fig:inclusive} our results are plotted as a function of the collider energy and compared with the combined experimental data from the {\sc Atlas} and {\sc Cms} collaborations.
In Table \ref{table:totalXS} our results at different collider energies are compared with corresponding results obtained by using the numerical code {\sc Top++}~\cite{Czakon:2011xx}, which implements the NNLO calculation of Refs.~\cite{Baernreuther:2012ws, Czakon:2012zr, Czakon:2012pz, Czakon:2013goa}, and we find excellent agreement. Our computation is performed with a value of the top mass $m_t=173.3$ GeV, considering \mbox{$n_F=5$} massless quark flavours and using the corresponding NNPDF31~\cite{Ball:2017nwa} sets of parton distribution functions with $\alpha_S(m_Z)=0.118$.
QCD scale uncertainties are evaluated using the standard 7-point scale variation, varying the factorisation ($\mu_F$) and renormalization ($\mu_R$) scales by a factor of $2$ around a common central value $\mu_0$ with the constraint $0.5\leq \mu_R/\mu_F\leq 2$. We choose as a central scale $\mu_0=m_t$.
\begin{table}[b]
\begin{center}
\begin{tabular}{|c|c|c|}
\hline
$\sigma_{NNLO}$
[pb] & $q_T$ subtraction & {\sc Top++} \\ \hline
8 TeV  & $238.5(2)^{+3.9\%}_{-6.3\%}$ & $238.6^{+4.0\%}_{-6.3\%}$ \\
13 TeV  & $794.0(8)^{+3.5\%}_{-5.7\%}$ & $794.0^{+3.5\%}_{-5.7\%}$ \\
100 TeV & $35215(74)^{+2.8\%}_{-4.7\%}$ & $35216^{+2.9\%}_{-4.8\%}$\\ \hline
\end{tabular}
\end{center}
\caption{
Total cross section for $t\bar{t}$ production in $pp$ collisions. The quoted uncertainties are obtained through scale variations as described in the text. Numerical uncertainties on the last digit are stated in brackets (including the $r_{cut}\to 0$ extrapolation uncertainties).
}
\label{table:totalXS}
\end{table}

We now present our result for distributions. We fix the center of mass energy to $\sqrt s = 13$ TeV and compare our results with the measurements from the {\sc Cms} collaboration~\cite{Sirunyan:2018wem}. To perform such comparison, no phase space cuts are applied, since the experimental results have been extrapolated to the inclusive phase space of parton level top quarks. Our predictions are multiplied by a factor of $0.438$, which corresponds to the semileptonic branching ratio of the top pair, and a factor of $2/3$ to take care of the fact that the measurement in Ref.~\cite{Sirunyan:2018wem} only considers the decay of two of the three lepton families.
An issue that in the differential case requires more care than in the inclusive one is the choice of the central scale $\mu_0$. It should be chosen of the order of the characteristic hard scale, which is different for each distribution: we choose the top mass $m_t$ for the rapidity distributions, the invariant mass of the $t \bar t$ pair $m_{t\bar t}$ for the invariant-mass distributions and the transverse mass $m_T$ for the transverse-momentum distributions.
We observed that the dynamic scale $H_T/2$ is of the same order as the characteristic scale for each distribution. We checked this statement by performing the computation with different scale choices: in the following only results for $\mu_0=H_T/2$ are shown.
The scale $H_T/2$ is defined as the average of the transverse masses of the top and anti-top quarks,
\begin{equation}
  \frac 12 H_T= \frac 12 \left(m_{T,t}+m_{T,\bar t}\right)\,,
\end{equation}
with
\begin{equation}
  m_{T,t(\bar t)}=\sqrt{m_t^2+p_{T,t(\bar t)}^2}\,.
\end{equation}

\begin{figure}[t]
  \begin{center}
    \includegraphics[width=0.25\textwidth]{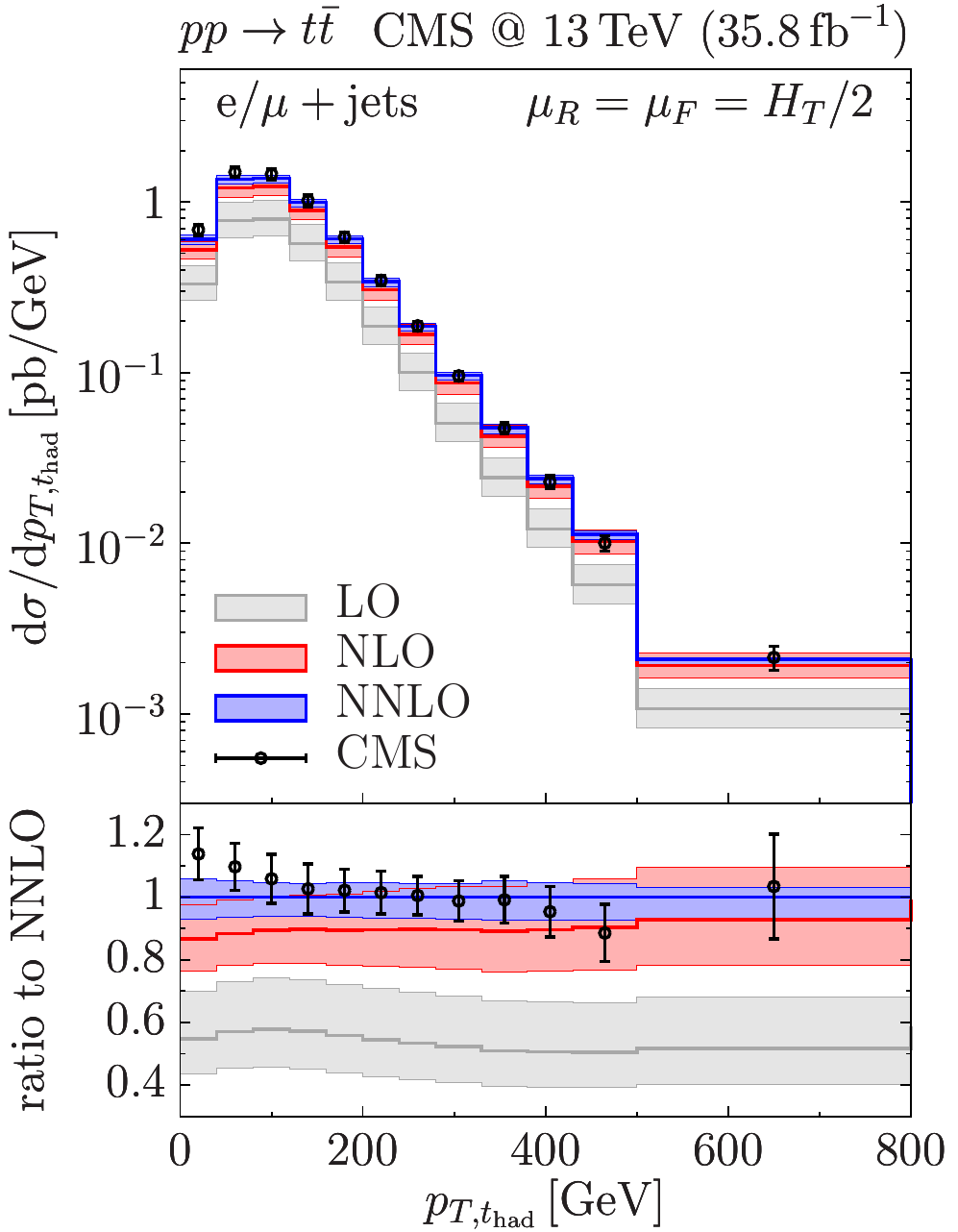}
    \includegraphics[width=0.25\textwidth]{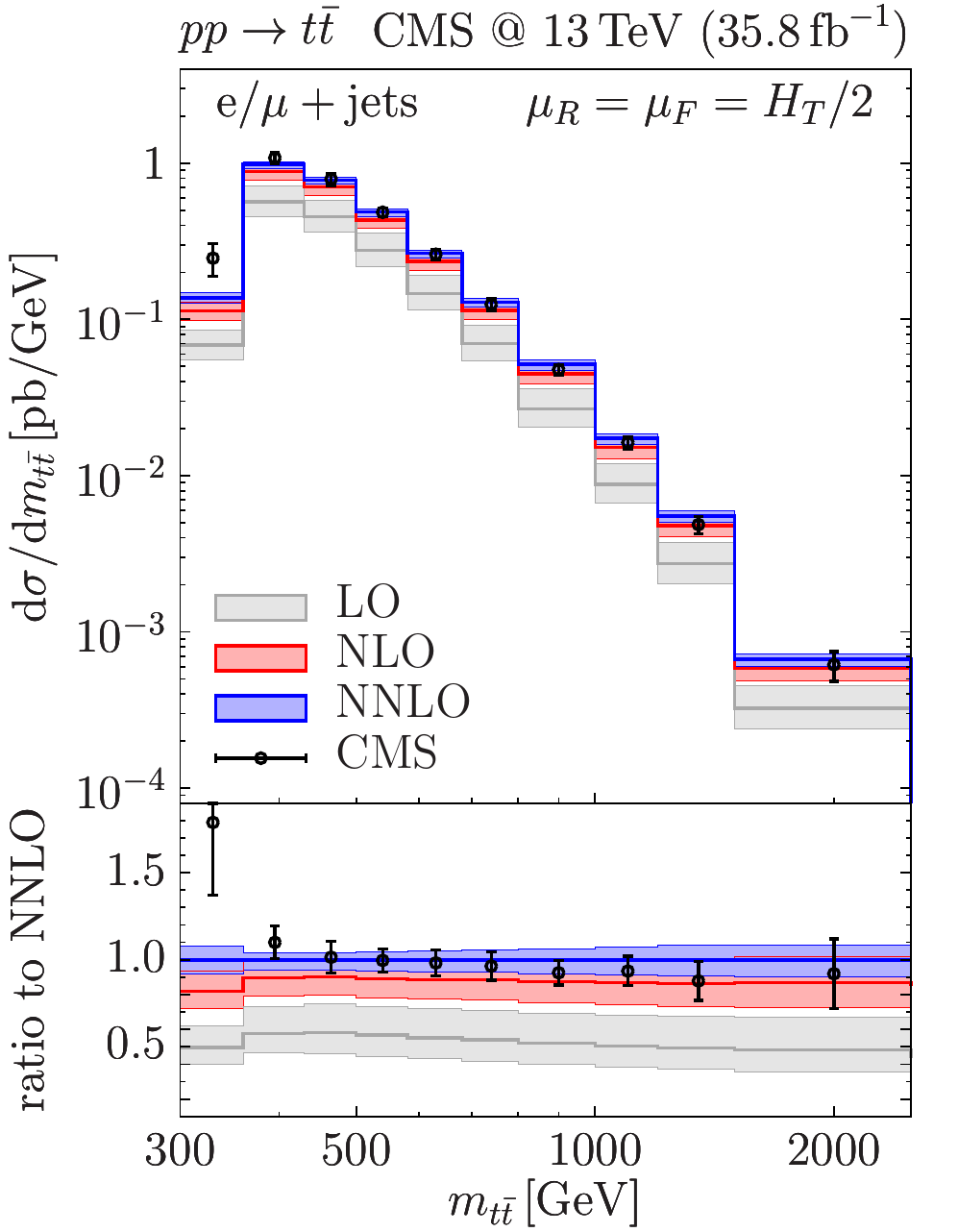}
    \includegraphics[width=0.25\textwidth]{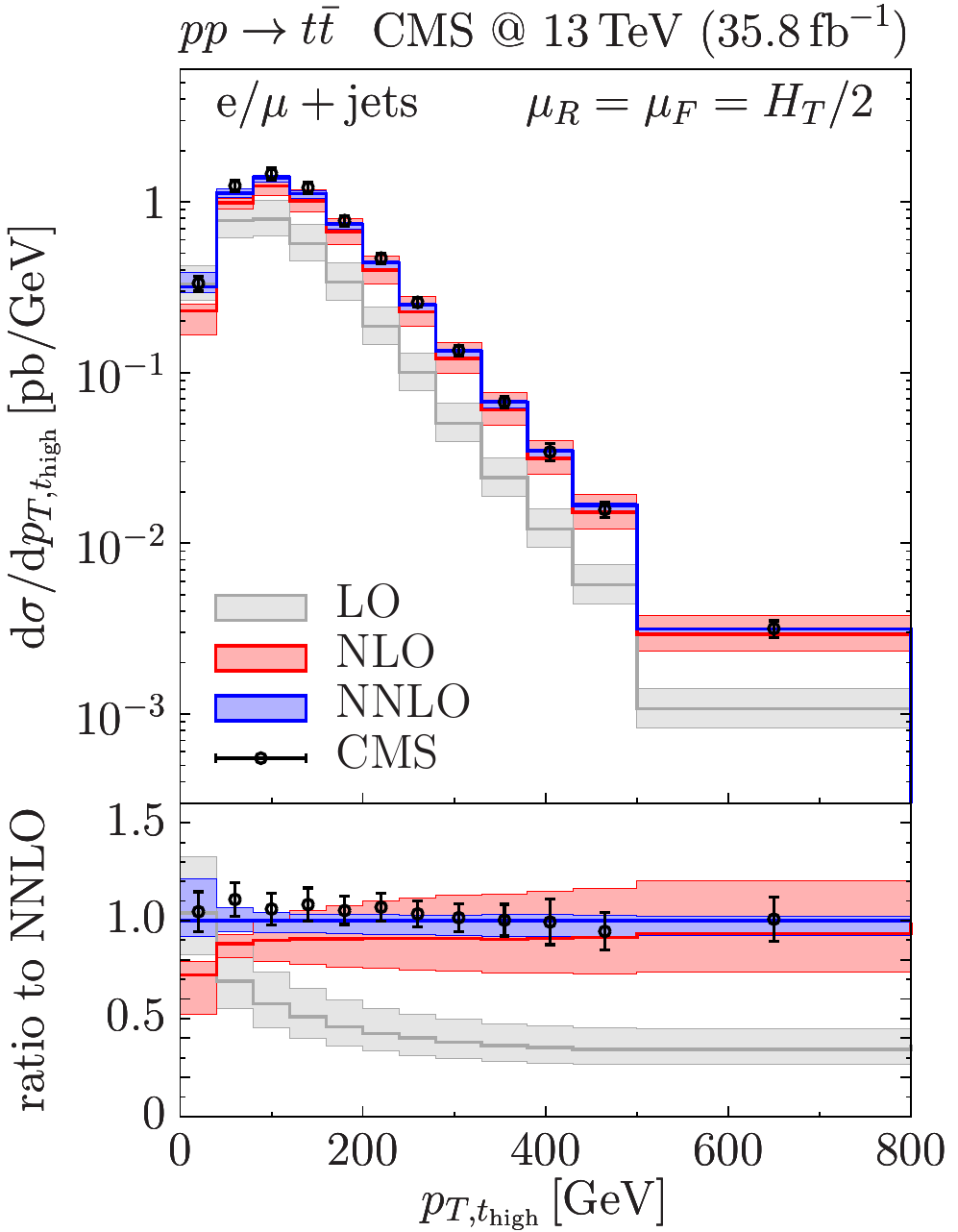}
    \includegraphics[width=0.25\textwidth]{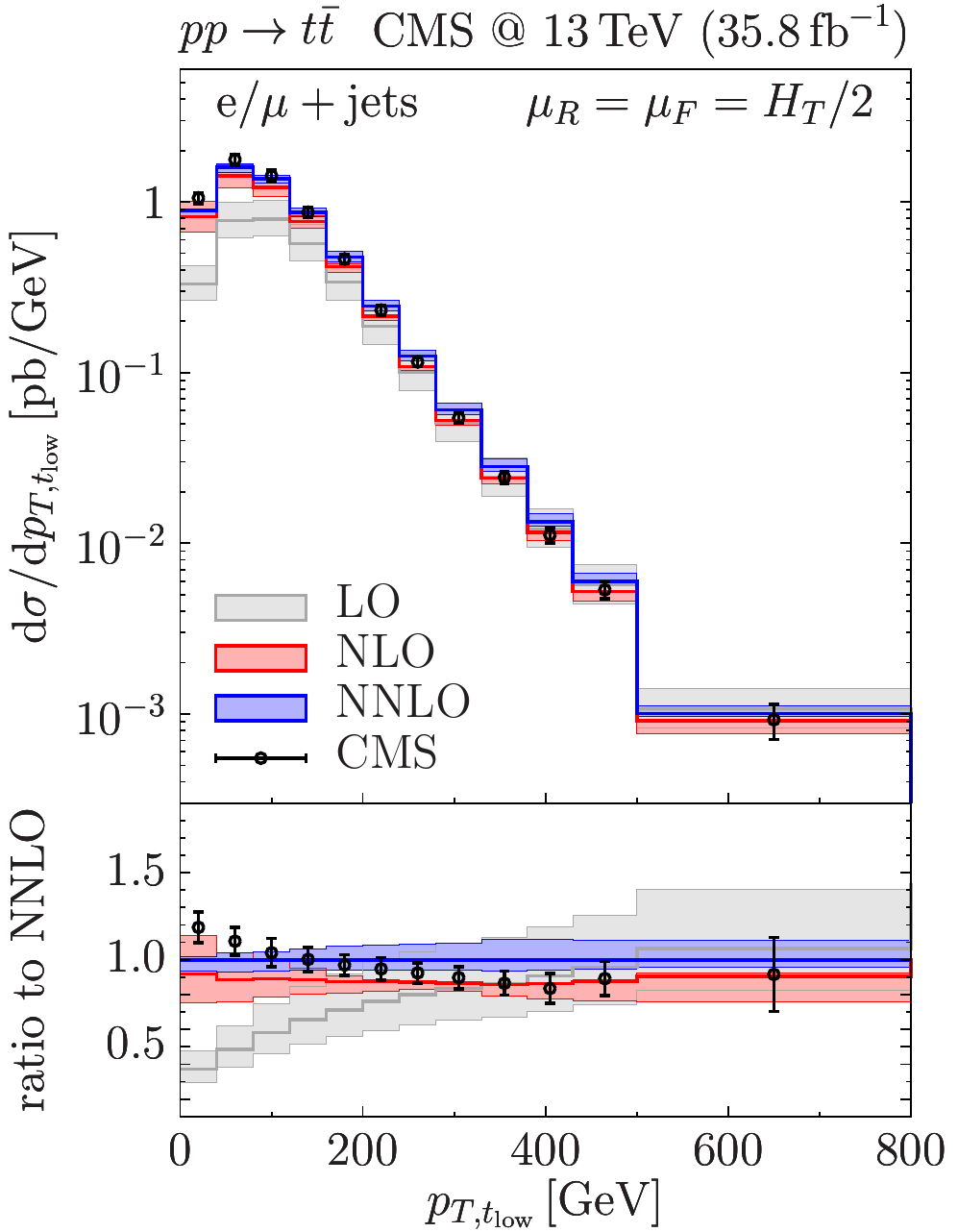}
  \end{center}
  \vspace{-0.5cm}
  \caption{From left to right: single-differential distributions as functions of $p_{T,\,t_{had}}$, $m_{t\bar t}$, $p_{T,\, t_{high}}$, $p_{T,\,t_{low}}$. LO, NLO and NNLO predictions are shown together with {\sc Cms} data~\cite{Sirunyan:2018wem}.}
  \label{fig:single_diff}
\end{figure}

We start by presenting our results for single differential distributions. Figure \ref{fig:single_diff} shows, in the upper panel of each plot, our results for the LO, NLO and NNLO predictions together with the data from the {\sc Cms} collaboration and, in the lower panels, the ratio of these quantities to our NNLO prediction.
As a general feature, one can observe that the first two plots (transverse momentum and invariant-mass distributions) do not present an overlap between LO and NLO bands, consistently with the results obtained in Table \ref{table:totalXS} for the inclusive cross section, while there is an overlap between NLO and NNLO bands, thus suggesting a convergence of the perturbative series. On the other hand, the last two plots present some instabilities in the shape of the distributions, which can be justified in terms of Sudakov-type divergences~\cite{Catani:1997xc}.

The first plot shows the transverse momentum distribution of the hadronically decaying top, $p_{T,\,had}$. Since we compute the cross section for stable top quarks, such prediction can be obtained by performing the average of the $p_T$ distribution of the top and anti-top quark. Data and theory are consistent within uncertainties, although the measured distribution is slightly softer than the theoretical prediction, as already pointed out in previous analyses of LHC data~\cite{Sirunyan:2018wem,Khachatryan:2015oqa,Khachatryan:2015fwh,Aad:2015hna,Aad:2015mbv,Khachatryan:2016gxp,Aaboud:2016iot,Aaboud:2016syx,Sirunyan:2017mzl}.

The second plot shows the invariant-mass distribution of the pair. Here we observe good agreement between data and theory, except for the first bin where the theoretical prediction by far underestimates the experimental measurement. There are several possible explanations for this disagreement: we first observe that this region, being the threshold region, is particularly sensitive to the value of the top-quark mass: a slightly lower value of the mass would make the prediction in this bin larger without significantly affecting the rest of the distribution. Besides, in this kinematical region the unfolding procedure~\cite{Sirunyan:2018wem} used to perform the conversion from the particle to the parton level becomes more delicate: both the value of the top mass and off-shell effect can have a large impact.

Finally, the last two plots show the $p_T$ distribution of the hardest and softest quark of the pair, respectively. As previously stated, these distributions present some shape instabilities. To justify them, let us start by considering the $p_{T,\, high}$ distribution in the $p_{T,\, high}\to0$ limit.
A small $p_{T,\, high}$ forces the transverse momentum of both quarks, and as a consequence the transverse momentum of the pair, to be small. We are thus exploring the small $p_{T,\, t\bar t}$ region, which is affected by Sudakov-type divergences at fixed order in QCD because of soft collinear emissions that unbalance the real and virtual contributions. Such a divergence does not appear in the final prediction because of the integration over the possible values of $p_{T,\, low}$ that smears out the unphysical fixed-order behaviour. The same effects appears in the $p_{T,\, low}$ distribution, but now the instability is spread over the full region of transverse momenta since $p_{T,\, low}\to0$ does not imply small $p_{T,\, high}$ anymore.
In both cases, accurate theoretical predictions in the entire $p_T$ region would require an all-order resummation of the logarithmically enhanced terms.

\begin{figure}[t]
  \begin{center}
    \includegraphics[width=0.65\textwidth]{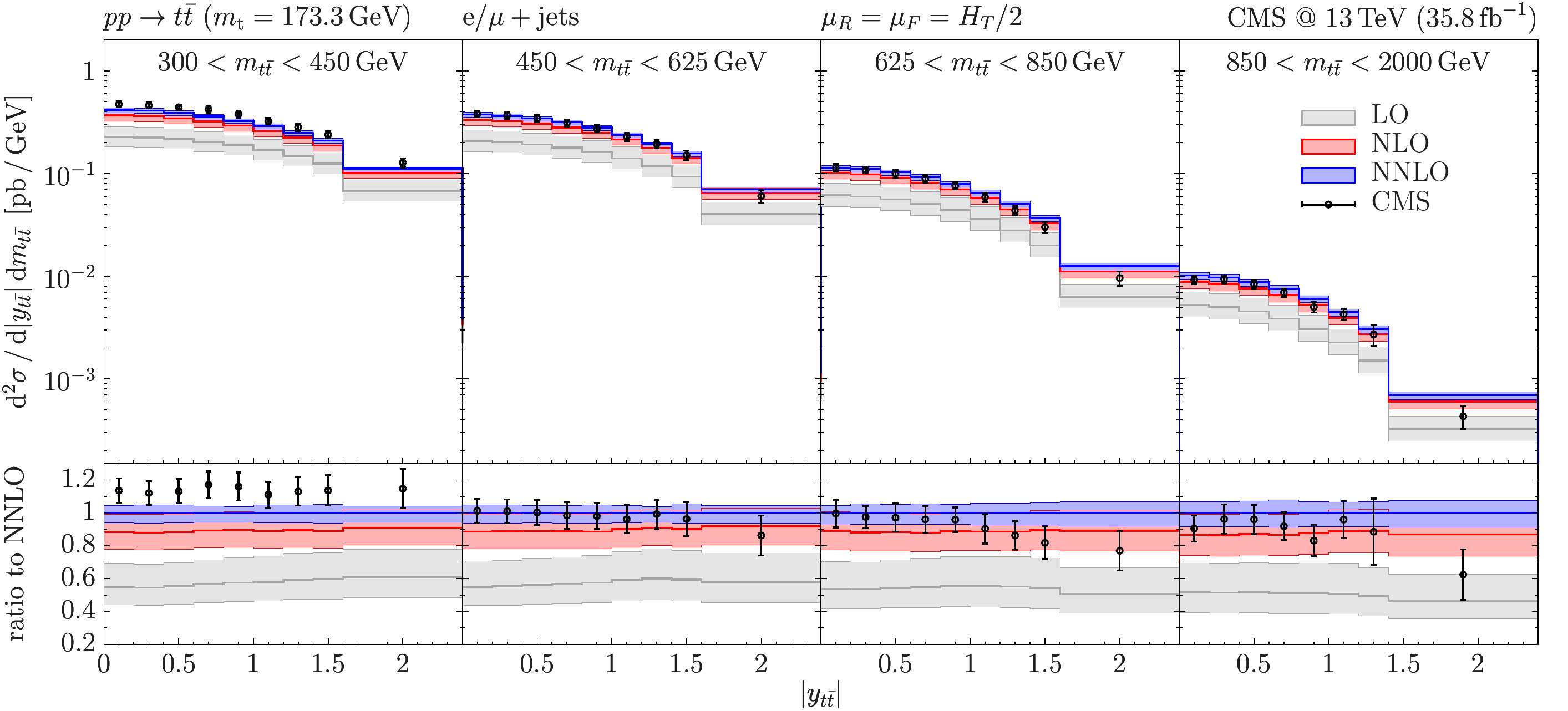}\\\vspace{0.5cm}
    \includegraphics[width=0.65\textwidth]{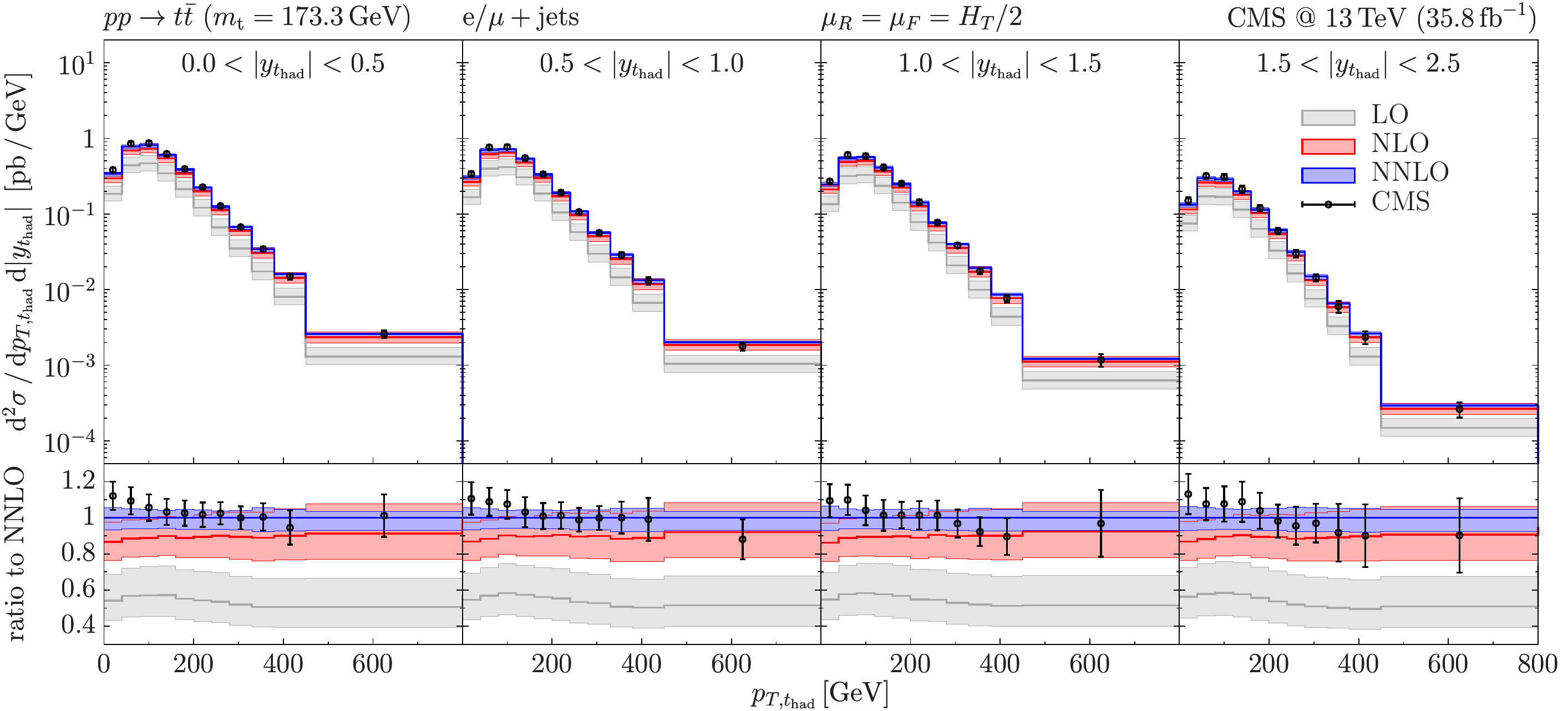}\\\vspace{0.5cm}
    \includegraphics[width=0.65\textwidth]{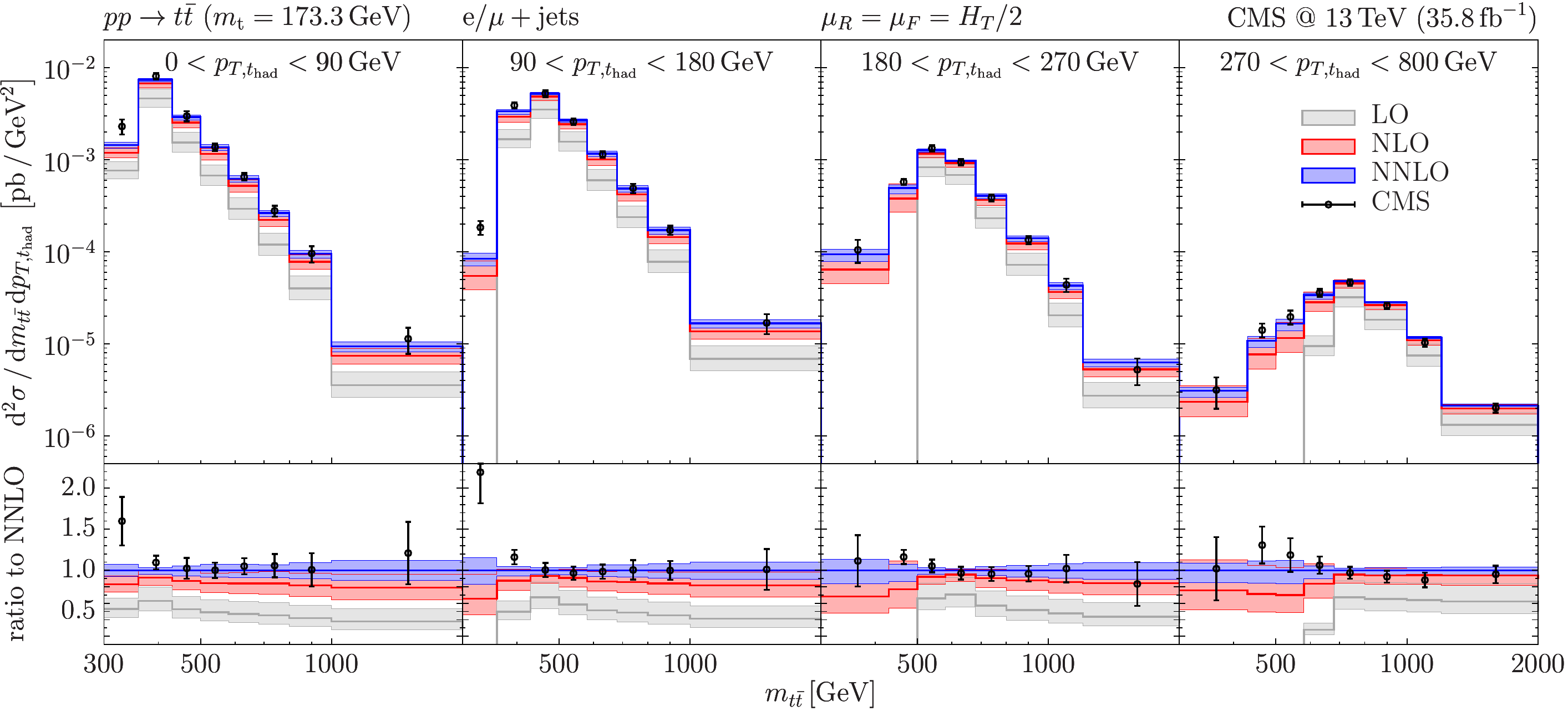}\\\vspace{0.5cm}
  \end{center}
  \vspace{-0.8cm}
  \caption{Double differential distribution as a function of $|y_{t\bar t}|$ in $m_{t\bar t}$ bins (top), as a function of $p_{T,\, t_{had}}$ in $|y_{had}|$ bins (middle), as a function of  $m_{t\bar t}$ in $p_{T,\, t_{had}}$ bins (bottom). LO, NLO and NNLO predictions are shown together with {\sc Cms} data~\cite{Sirunyan:2018wem}.}
  \label{fig:double_diff}
\end{figure}

We now consider the double-differential distributions. Figure \ref{fig:double_diff} shows again, in the upper panel of each plot, our theoretical prediction together with the data from the {\sc Cms} collaboration, while the lower panels the ratio of the same quantities to our NNLO prediction.

The first plot shows the rapidity distribution in invariant-mass bins. Also in the case of the rapidity distributions one can observe that LO and NLO bands do not overlap, while NLO and NNLO do overlap. Besides, the impact of radiative corrections appears to be uniform in both variables. Considering the agreement with the data, we notice that, in the first invariant mass bin, the experimental measurement overshoots the theoretical prediction, consistently with what we observed in the single-differential distribution as a function of the invariant mass. In this case, the effect is not as sizeable because of a larger size of the first invariant-mass interval.

The second plot shows the transverse momentum distribution of the hadronically decaying top quark in rapidity bins. In each rapidity bin, one observes the same features we already noticed in the single-differential distributions: overlap of the NLO and NNLO bands and the data being softer than the NNLO prediction.

Finally, the third plot shows the invariant-mass distribution in $p_{T,\, had}$ intervals. These distributions are affected by a kinematical boundary at LO, namely $m_{t\bar t}>2m_{T,\,min}$. As a consequence, below this threshold the (N)NLO is effectively (N)LO which implies larger scale uncertainties. Above this cut, the LO distribution sharply increases, causing shape instabilities due to soft-collinear radiation. Nevertheless, one still observes an overall good agreement between data and theory except for the threshold region.

\section{Conclusions}
We have completed a fully differential NNLO computation of top pair production at hadron colliders by using the $q_T$-subtraction formalism. This constitutes the first complete application of the $q_T$ subtraction formalism to the production of a colourful final state at NNLO. The process has been implemented in the {\sc Matrix} framework: our code is now able to perform fast and efficient computations of fiducial cross section and multi-differential distributions.
We compared our results with those available in the literature, finding excellent agreement, and with recent measurements from the {\sc Cms} collaboration.

Our calculation will be made public in a future release of the {\sc Matrix} program, providing a flexible tool to evaluate multi-differential distributions with arbitrary cuts on the top-quark kinematic variables.

\bibliography{biblio}

\end{document}